\documentclass[%
 reprint,
superscriptaddress,
nofootinbib,
 amsmath,amssymb,
 aps,
prc,
]{revtex4-2}

\usepackage[compatibility=false]{caption}
\usepackage{subcaption}

\usepackage{graphicx}
\usepackage{dcolumn}
\usepackage{bm}

\usepackage{booktabs}
\usepackage{xcolor,colortbl}
\usepackage{hyperref}

\begin{document}

\preprint{APS/123-QED}

\title{Particle spectra in the integrated hydrokinetic model at RHIC Beam-Energy-Scan energies}


\author{Narendra Rathod}
\affiliation{%
 Warsaw University of Technology, Faculty of Physics, Koszykowa 75, 00-662 Warsaw, Poland\\
}%
\affiliation{%
University of Bern, DBMR and University Clinic for Nuclear Medicine, Inselspital Bern, Switzerland\\
}%

\author{Yuri Sinyukov}%
\affiliation{%
Bogolyubov Institute for Theoretical Physics,
Metrolohichna  14b, 03143 Kyiv,  Ukraine 
}%
\affiliation{%
 Warsaw University of Technology, Faculty of Physics, Koszykowa 75, 00-662 Warsaw, Poland\\
}%

\author{Musfer Adzhymambetov}
\email{adzhymambetov@gmail.com}
\affiliation{%
Bogolyubov Institute for Theoretical Physics,
Metrolohichna  14b, 03143 Kyiv,  Ukraine 
}%

\author{Hanna Zbroszczyk}
\affiliation{%
 Warsaw University of Technology, Faculty of Physics, Koszykowa 75, 00-662 Warsaw, Poland\\
}%


\date{\today}

\begin{abstract}
We study light-hadron production in Au+Au collisions at $\sqrt{s_{NN}} = 7.7–39$ GeV using an extended Integrated HydroKinetic Model (iHKMe). Focusing on transverse momentum spectra, we investigate the sensitivity to key model parameters, particularly the thermalization timescale. We consider two distinct equations of state: one featuring a crossover and the other a first-order phase transition. In both cases, thermalization begins shortly before full nuclear overlap and lasts approximately 1~fm/$c$ across all energies. Both equations of state provide a similarly good description of the soft particle momentum spectra once the other parameters are slightly adjusted. The most pronounced differences arise at the lower RHIC BES energy of $\sqrt{s_{NN}} = 7.7$~GeV, particularly in proton and kaon yields, reflecting their sensitivity to the freeze-out parameters.

\end{abstract}

\maketitle

\section{Introduction}
Relativistic heavy-ion collisions, such as those at the Large Hadron Collider (LHC) and the Relativistic Heavy Ion Collider (RHIC), are known to create new forms of strongly interacting matter. At extreme temperatures and densities, quasi-free quarks and gluons can form a nearly equilibrated state known as the quark–gluon plasma (QGP) \cite{Shuryak:1980tp}. In contrast, under ordinary terrestrial conditions, quarks and gluons remain confined within color-neutral hadrons \cite{Karsch:2001cy, Aoki:2006we}.

In the quantum chromodynamics (QCD) phase diagram, lattice QCD calculations indicate that the quark–gluon plasma and hadron–resonance gas (HRG) phases are connected by a smooth crossover transition at vanishing baryon chemical potential \cite{Aoki:2006we, Bazavov:2011nk, Ding:2015ona}, a situation realized at LHC and top RHIC energies. However, at higher baryon densities, the system is expected to undergo a genuine phase transition, potentially of first order \cite{Stephanov:2004wx, Fodor:2004nz, Bzdak:2019pkr}. If such a phase transition line exists in the QCD phase diagram, it must end at a critical point, beyond which the transition becomes a crossover. Ongoing and upcoming experiments \cite{STAR:2010vob, CBM:2016kpk, NA61:2014lfx} at lower collision energies, typically in the range of several GeV per nucleon pair, aim to identify clear signals of this phase transition and locate the critical endpoint.

With the launch of the Beam Energy Scan (BES) program at RHIC, three primary objectives were outlined  \cite{Odyniec:2013kna, STAR:2014egu}. The first goal is to explore the QCD phase diagram by varying the center-of-mass collision energy, $\sqrt{s_{NN}}$, in the range from 7.7 to 39 GeV. This variation enables access to different regions of temperature $T$ and baryon chemical potentials $\mu_B$ during the system's evolution, facilitating the search for the onset of deconfinement. The second goal is to investigate fluctuations and enhanced susceptibilities that are expected to increase near the QCD critical point \cite{Stephanov:1998dy, Stephanov:1999zu, Hohenberg:1977ym}. However, such critical signatures may be suppressed due to conservation laws and the finite size and lifetime of the system. This challenge leads to the third goal: to identify observables sensitive to the softening of the QCD equation of state (EoS), which may signal the presence of a first-order phase transition. 

One of the main challenges in analyzing and interpreting experimental results from heavy-ion collisions stems from the complex and dynamic evolution of the system, combined with its small spatial extent and extremely short lifetime. The properties of strongly interacting matter, particularly its behavior near a potential phase transition or critical point, affect experimental observables only indirectly. To extract these effects quantitatively, or even estimate them qualitatively, it is often necessary to construct a comprehensive dynamical model that simulates the system’s entire evolution, from the initial impact to the final detection of produced hadrons.

Relativistic hydrodynamics has emerged as the most effective framework for describing the collective expansion of the strongly interacting matter created in these collisions. However, it comes with notable limitations: it is primarily applicable to soft observables (low transverse momentum, $p_T$) and assumes local thermal equilibrium \cite{Romatschke:2017ejr}. To address these constraints, theorists have developed hybrid models that integrate relativistic hydrodynamics, used for the hot, dense equilibrated stage, with microscopic transport models that describe out-of-equilibrium dynamics. Such non-equilibrium conditions are expected both in the very early pre-equilibrium stage, before thermalization occurs, and in the late afterburner stage, when the system becomes dilute and departs from equilibrium due to expansion. 

At the LHC and the highest RHIC energies, the standard model of heavy-ion collisions is well established and widely accepted \cite{Romatschke:2009im, Heinz:2013th, Huovinen:2006jp, Gale:2013da}. This model typically divides the collision process into several distinct stages: the formation of a non-equilibrium state during the penetration of the colliding nuclei, subsequent thermalization process leading to formation of QGP, hydrodynamic expansion of the resulting fireball, hadronization of the QGP into HRG, and finally, the hadronic cascade stage \cite{Dumitru:1999sf, Lin:2004en, Petersen:2008dd, Schenke:2010nt, Steinheimer:2011mp,  Gale:2013da, Werner:2013tya, Shen:2014vra, Naboka:2015qra}. As the collision energy decreases, this overall picture remains qualitatively similar but becomes significantly more complex. First, the characteristic time for two gold nuclei to fully overlap is approximately  $10^{-3}$~fm/$c$ at $\sqrt{s_{NN}}=5.02$ TeV, while at $\sqrt{s_{NN}}=14.5$ GeV at BES RHIC it exceeds 1.5~fm/$c$. Consequently, the system becomes highly inhomogeneous not only in the transverse but also in the longitudinal direction, leading to a significant violation of boost invariance (i.e.\ a breakdown of the Bjorken picture \cite{Bjorken:1982qr}). Second, the lower energy density leads to a slower thermalization process, which may significantly shorten, or even eliminate, the hydrodynamic phase. At collision energies of just a few GeV, it is plausible that the pre-equilibrium stage dominates the evolution, preventing the system from ever reaching full thermal equilibrium. 

In a recent study \cite{Adzhymambetov:2024zzz}, we introduced an extended framework to describe the complete space–time evolution of the system created in relativistic heavy-ion collisions at intermediate and low energies, from the initial nuclear overlap to the final decoupling of free-streaming hadrons. This extended version of the integrated hydrokinetic model, denoted iHKM\textsubscript{e}, employs relativistic hydrodynamics for the bulk equilibrated medium evolution, while the UrQMD cascade \cite{Bleicher:1999xi} is utilized to account for the essential non-equilibrium dynamics. Unlike many hybrid approaches \cite{Steinheimer:2012bn, Shen:2017ruz, Akamatsu:2018olk, Wu:2021fjf}, which segment the evolution into distinct phases in space-time, iHKMe integrates microscopic and macroscopic descriptions concurrently. In our approach, the system initially exists far from equilibrium and gradually thermalizes, transitioning into a hydrodynamic regime\footnote{Other approaches to a continuous transition from UrQMD to hydrodynamics are discussed in \cite{Akamatsu:2018olk, Du:2018mpf}}. Since the thermalization rate is not well constrained, it is treated as a free parameter in the model. By tuning this parameter, iHKMe enables a smooth interpolation between scenarios dominated by rapid thermalization (as at LHC energies) and those that remain largely non-equilibrated throughout the evolution, as expected for very low-energy collisions.

This study continues our previous work \cite{Adzhymambetov:2024zzz}, in which we introduced an extended version of the integrated hydrokinetic model. In the present paper, we focus on describing the transverse momentum spectra of particles produced in Au-Au collisions at $\sqrt{s_{NN}}= 7.7- 39$~GeV, as part of the RHIC Beam Energy Scan program. We calibrate the model’s free parameters using two different equations of state: one featuring a smooth crossover transition \cite{Steinheimer:2010ib}, and the other incorporating a first-order phase transition with a soft equation of state \cite{Kolb:2003dz}.

\section{Model}

The iHKM model was originally developed to study ultrarelativistic heavy-ion collisions \cite{Naboka:2015qra}. We have recently extended the model to accommodate intermediate and low collision energies, broadening its applicability down to a few GeV. A detailed description of the extended model can be found in \cite{Adzhymambetov:2024zzz}. In this paper, we provide only a brief overview of the model's five evolutionary stages, discuss the potential overlap of these stages at lower collision energies, and introduce the key free parameters that govern the model's dynamics. 

\subsubsection{Initial pre-equilibrium dynamics}

One of the key differences between ultrarelativistic heavy-ion collisions at the LHC and the lower-energy experiments of the RHIC BES program lies in the very different relative velocities of colliding nuclei (and correspondingly space-time Lorentz factors for them), which, according to QCD approaches, leads to different physical pictures of nucleus-nucleus collisions at different energies. First, it concerns the different initial states for the colliding process. At TeV-scale energies, partons from the two nuclei interact only during the extremely short overlap time, approximately $10^{-3}$ fm/$c$, which is negligible compared to the full evolution timescale of the system (typically exceeding 10 fm/$c$). In contrast, at lower energies, the scenario becomes significantly more complex. The duration of the nuclear overlap, which increases with decreasing beam energy, can be estimated using the following equation:
\begin{equation}\label{eq:overlapTime}
    \tau_{\text{overlap}} = \frac{2 R_N}{\sqrt{\left(\sqrt{s_{NN}} / 2 m_{N}\right)^{2} - 1}},
\end{equation}
where $R_N$ is the radius of a nucleus and $m_N$ is the nucleon mass. Here we adopt the proper time $\tau$ and space-time rapidity $\eta$ variables, following the conventional Bjorken coordinate system for longitudinally expanding systems \cite{Bjorken:1982qr}.  A straightforward calculation shows that at intermediate energies, around  $\sqrt{s_{NN}} \sim 10$~GeV, the overlap time already exceeds $1.5$~fm/$c$. Thermalization only begins during this stage, meaning that a purely hydrodynamic description is not yet applicable. Consequently, a microscopic approach must be employed to model the system's early, out-of-equilibrium dynamics.

In the iHKM for ultra-relativistic energies, the initial state of matter is generated using the Monte Carlo Glauber model, implemented in the GLISSANDO II package \cite{Rybczynski:2013yba}, featuring a longitudinally boost-invariant distribution in coordinate space and a highly anisotropic momentum distribution inspired by the Color Glass Condensate framework \cite{McLerran:1994vd, Iancu:2003xm, Gelis:2010nm, Florkowski:2010cf}. In the current version, event-by-event UrQMD cascade simulations are employed to describe the initial non-equilibrium dynamics up to a proper time $\tau_0$, which marks the onset of the thermalization stage, discussed in detail later. 

The output of UrQMD simulations is used to construct the stress-energy tensor and baryon charge current, which serve as initial conditions for the hydrodynamic equations \cite{Petersen:2008dd, Huovinen:2009yb, Oliinychenko:2016vkg}. At each step of proper time $\tau_j \pm \Delta \tau/2$, we select all particle tracks\footnote{Here we refer to the particle positions at each step of Cartesian time $t$ obtained from UrQMD. For consistency, the same time step is used in UrQMD and in the subsequent hydrodynamic simulations, $\Delta t = \Delta \tau$.
 } (labeled by index $i$) that satisfy
\begin{equation}\label{eq:timeselection}
    \left|\sqrt{t_i^2 - z_i^2} - \tau_j \right| < \frac{\Delta \tau}{2}\,.
\end{equation}
The space-time distributions of the macroscopic fields are then constructed as
\begin{align}\label{eq:urqmdTensors}
    T^{\mu\nu}_{\text{urqmd}} (x_j) &= \sum_{i} \frac{p_i^{\mu} p_i^{\nu}}{p_i^{0}} \, {\cal K}_{ij},\\
    J^{\mu}_{\text{urqmd}}(x_j) &= \sum_{i} B_i \frac{p_i^{\mu}}{p_i^{0}} \, {\cal K}_{ij}.
\end{align}
Here, $T^{\mu\nu}_{\text{urqmd}}$ represents the stress-energy tensor at the space-time point $x^{\mu}_j$, $J^{\mu}_{\text{urqmd}}(x_j)$ is the baryon current\footnote{In this paper, we do not consider separate equations for other conserved charges.
The electric charge density is assumed to be proportional to the baryon density, $n_q = \frac{Z}{A} n_b$, where $\frac{Z}{A} \approx 0.4$ for gold nuclei. The strangeness is always locally zero. }, $p^{\mu}_i$ denotes the four-momentum of particle $i$, and $B_i$ represents its baryon charge. We also utilize a  ${\cal K}_{ij}$ for particle smearing to generate a relatively smooth or averaged over the ensemble of similar events tensors.
\begin{equation}\label{eq:kernelIHKM}
{\cal K}_{ij}= \frac{n_{j}^{\lambda}  u_{i
\lambda} }{(\pi R^2)^{3/2}} \exp\left( \frac{ r_{ij}^{\mu} \left( g_{\mu\nu} - u^{i}_{\mu}u^{i}_{\nu} \right) r_{ij}^{\nu} }{R^2}\right).
\end{equation}
In this expression, $r_{ij}^{\mu} = x^{\mu}_{i} - x^{\mu}_j$ represents the four-vector connecting the particle and the center of the cell ($x_i, x_j \in \sigma^{\mu}$) in $3+1$-dimensional spacetime,
 $u_{i\lambda}$ is four-velocity of particle, and $n_{j}^{\lambda}$ is a normal vector to the hypersurface on the lattice $\Delta \sigma^{\mu}$
 
\begin{align}\label{eq:milneNormal}
    \Delta \sigma^{\mu} &=  n^{\mu} \Delta x \Delta y \Delta \eta, \\
    n^{\mu} &=  \left(\tau \cosh \eta, 0, 0, \tau \sinh \eta\right)\,,
\end{align}

Here, $\Delta x = \Delta y = 0.3$~fm denote the cell size in the transverse plane, and $\Delta \eta = 0.05$ is the cell size of the space-time rapidity in the longitudinal direction. We use the same lattice spacing, as well as the proper time step $\Delta \tau = 0.05$~fm/$c$, for all energies throughout this paper.

In Eq.~(\ref{eq:kernelIHKM}), a free scalar parameter of the model $R$ is introduced. It is responsible for the smoothness of the initial collisions. Lastly, let us specify that $\sum_j K_{ij} \approx 1$, ensuring no noticeable violations of conservation laws in this procedure~\cite{Adzhymambetov:2024zzz}.

\subsubsection{Thermalzaition}
During the thermalization stage, the system is assumed to consist of two coexisting phases. The first is a non-equilibrium component that evolves according to the UrQMD model, as described by Eq.~(\ref{eq:urqmdTensors}). The second corresponds to near-equilibrium matter, which is governed by relativistic hydrodynamics. Its macroscopic tensors are given by
\begin{equation}\label{Ttilde}
    T^{\mu\nu}_{\text{hydro}}(x) = (\epsilon + p) u^{\mu} u^{\nu} - p g^{\mu\nu} + \pi^{\mu\nu},
\end{equation}
\begin{equation}\label{eq:Jtilde}
    J^{\mu}_{\text{hydro}}(x) = n_B u^{\mu},
\end{equation}
where $\epsilon$, $p$, and $n_B$ denote the local energy density, pressure, and baryon density, respectively; $u^{\mu}$ is the fluid four-velocity, $g^{\mu\nu}$ is the metric tensor, and $\pi^{\mu\nu}$ is the shear-stress tensor. In this work, we neglect other dissipative effects such as bulk viscosity, heat flow, and baryon diffusion, focusing solely on the shear-stress contribution in the hydrodynamic evolution. The pressure and energy density are related through the equation of state.

A linear combination of two components gives the total tensors of the system:
\begin{equation}\label{eq:tensorSplit1}
    T^{\mu\nu}_{\text{total}}(x) = T^{\mu\nu}_{\text{urqmd}}(x) \cdot {\cal P}_{\tau} + T^{\mu\nu}_{\text{hydro}}(x)\cdot \left( 1- {\cal P}_{\tau} \right),
\end{equation}
\begin{equation}\label{eq:tensorSplit2}
    J^{\mu}_{\text{total}}(x) = J^{\mu}_{\text{urqmd}}(x) \cdot {\cal P}_{\tau}+ J^{\mu}_{\text{hydro}}(x)\cdot \left( 1- {\cal P}_{\tau} \right),
\end{equation}
where ${\cal P}_{\tau} = {\cal P}(\tau)$ is a weight function such that ${\cal P}(\tau_0)=1$ at the start of the thermalization stage, ${\cal P}(\tau_{\text{th}})=0$ at the end, and $0<{\cal P}(\tau_0<\tau<\tau_{\text{th}})<1$ in between. So this weight function regulates how matter is pumped from one component to another during the thermalization. As in our previous papers \cite{Akkelin:2009nz, Naboka:2014eha}, we utilize an ansatz for ${\cal P}(\tau)$ inspired by the Boltzmann equation in relaxation time approximation.
\begin{equation}\label{eq:Ptau}
    {\cal P}(\tau ) =  \left(\frac{\tau_{\text{th}}-\tau}{\tau_{\text{th}}-\tau_0} \right)^{\frac{\tau_{\text{th}}-\tau_{0}}{\tau_{\text{rel}}}}.
\end{equation}
This function involves three free parameters: the onset of thermalization $\tau_0$, the end of thermalization $\tau_{\text{th}}$, and its rate encoded in the relaxation time $\tau_{\text{rel}}$.

\subsubsection{Hydrodynamic expansion, particlization, and hadron cascade}
After thermalization is achieved, iHKM follows a classical path of hybrid models. At $\tau_{th}$, the system expands via viscous hydrodynamic equations, which we solve numerically using the vHLLE code \cite{Karpenko:2013wva}. Once the system becomes dilute and reaches the critical energy density $\epsilon_{\text{sw}}$ \footnote{As in our previous work \cite{Adzhymambetov:2024zzz}, we use the energy density obtained from the hydrodynamic tensor $T^{\mu\nu}_{\rm hydro}$ according to Eq.~(\ref{Ttilde}).}, the evolution switches back to the UrQMD cascade. To determine the particlization hypersurface of constant energy density, we employ the Cornelius routine \cite{Huovinen:2012is, Molnar:2014fva}. 
The particlization of liquid into hadron resonant gas is realized via a Cooper-Frye-like algorithm 
\begin{equation}\label{CooperFrye}
N_{i} = \int d^3p \int \frac{ d \sigma_{\mu} p^{\mu}}{p_0} f_i(x, p).
\end{equation}
Here $N^{i}$ is a number of hadrons of species $i$ emitted from the element of hypersurface $d \sigma_{\mu}$ and $f_{i}(x, p)$ is a distribution function.
For high-energy collisions, the Bose-Einstein or Fermi-Dirac equilibrium distribution function with a small correction is a fairly good assumption. However, if the equilibrium is not reached, which we expect at several GeV experiments, two components must be considered. Following Eqs.~(\ref{eq:tensorSplit1}), (\ref{eq:tensorSplit2}) in relaxation time approximation we obtain 
\begin{equation}\label{eq:dist_function}
    f(x, p) = \left( 1 - {\cal P}(\tau) \right)f_{\texttt{eq.}}(x, p) + {\cal P}(\tau)f_{\texttt{n.eq.}}(x, p), 
\end{equation}
where $f_{\texttt{eq.}}(x, p)$ is a near-equilibrium component constructed from the output of the hydrodynamic component, and $f_{\texttt{n.eq.}}(x, p)$ is a far from equilibrium distribution function constructed from UrQMD with a kernel (\ref{eq:kernelIHKM}). Consequently, we sampled hadrons for this component from the same UrQMD event that was used for initial dynamics.

\section{Finetuning and results}
\subsubsection{Free parameters}
To assess the sensitivity to the softening of the equation of state, we consider two qualitatively distinct models: the chiral EoS \cite{Steinheimer:2010ib}, which exhibits a crossover (CO) transition between the quark-gluon plasma and the hadron resonance gas, and an EoS featuring a first-order phase transition (PT) \cite{Kolb:2003dz}. For each EoS, we conduct simulations while systematically varying the following free parameters introduced in the previous section:

\begin{itemize}
\item $\tau_{0}$ $\left[\text{fm}/c\right]$ - start of the thermalization stage
\item $\eta/s$ $\left[1\right]$ - shear viscosity to entropy density ratio. Scaling factor in shear stress tensor $\pi^{\mu\nu}$ 
\item $\tau_{\text{th}}$ $\left[\text{fm}/c\right]$ - end of thermalization
\item $\epsilon_{\text{sw}}$ $\left[\text{GeV/fm}^3\right]$ - hydrodynamic energy density corresponding to the transition to the hadron afterburner stage
\item $\tau_{\text{rel}}$ $\left[\text{fm}/c\right]$ - relaxation time
\item $R$ $\left[\text{fm}\right]$  - gaussian smearing parameter in kernel~(\ref{eq:kernelIHKM})
\end{itemize}

To simplify the calibration procedure and reduce the number of free parameters, we fix the relaxation time to its maximum permitted value, effectively transferring all remaining flexibility to the $\tau_{\text{th}}$ parameter.

\begin{equation}\label{eq:tauth}
\tau_{\text{rel}} = \tau_{\text{th}} - \tau_0.
\end{equation}

\subsubsection{Sensitivity of spectra to the parameters}

To investigate correlations between model parameters and observables, we simulate 750 sets of Au--Au collisions at $\sqrt{s_{NN}} = 14.5$~GeV within the $20$--$30\%$ centrality class. Parameter values for each simulation are sampled uniformly using Latin hypercube sampling within the ranges specified in Table~\ref{tab:latinCube}. In this analysis, both the $R$ and $\tau_{\rm rel}$ parameters are varied, while ensuring that the constraint on $\tau_{\rm th}$ given by Eq.~(\ref{eq:tauth}) is satisfied.

In the RHIC BES experiments, centrality classes are determined by dividing minimum-bias events according to the charged-particle multiplicity at midrapidity \cite{STAR:2019vcp}. However, simulating minimum-bias events for each parameter set is computationally expensive, since it requires performing a large number of full hydrodynamic runs. Instead, we determine centrality using multiplicities from the UrQMD events that serve as initial conditions for iHKM. For each collision energy, we simulate a large ensemble of UrQMD events up to $t = 200$~fm/$c$, with the maximum impact parameter set to $13.2$~fm, corresponding to approximately the $0$--$80\%$ centrality range according to a Glauber model with an inelastic NN cross section of 30~mb \cite{Eskola:1988yh, STAR:2012och}. The charged-particle multiplicity at midrapidity is evaluated for each event, and the ensemble is subdivided into narrower centrality classes. Assuming a correlation between final multiplicity in full iHKM runs and that in the corresponding UrQMD events, each event can be assigned to a centrality class before the hydrodynamic stage.

\begin{table}[]
    \centering
    \begin{tabular}{|l|c|c|}
        \hline
        \textbf{Parameter} & \textbf{Min Value} & \textbf{Max Value} \\
        \hline
        $\tau_0$ (fm/$c$) & 0.8 & 2.2 \\
        \hline
        $\tau_{\rm rel}$ (fm/$c$) & 0.3 & 2.0 \\
        \hline
        $\eta/s$ & 0.0 & 0.2 \\
        \hline 
        $\epsilon_{\rm sw}$ (GeV/fm$^3$) & 0.35 & 0.55 \\
        \hline
        $R$ (fm) & 0.4 & 1.2 \\
        \hline
    \end{tabular}
    \caption{Ranges of model parameters used in the Latin hypercube sampling.}
    \label{tab:latinCube}
\end{table}

\begin{figure*} 
\centering
\includegraphics[width=0.32\textwidth,height=0.22\textheight]{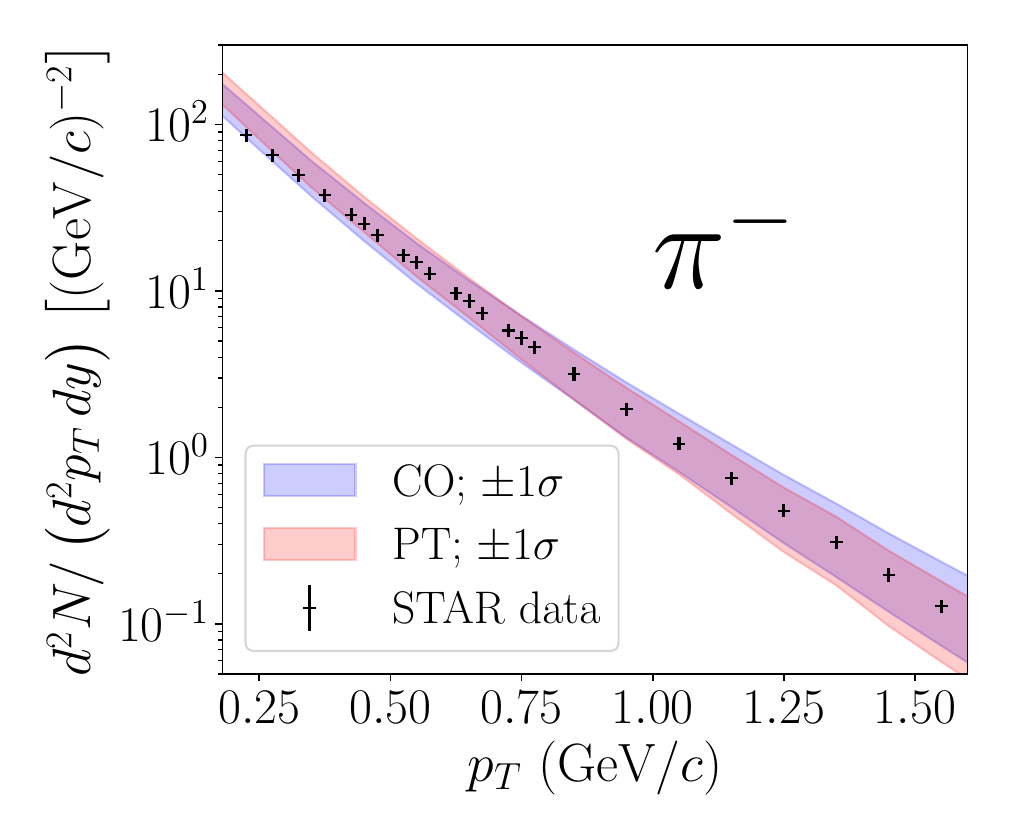}
\includegraphics[width=0.32\textwidth,height=0.22\textheight]{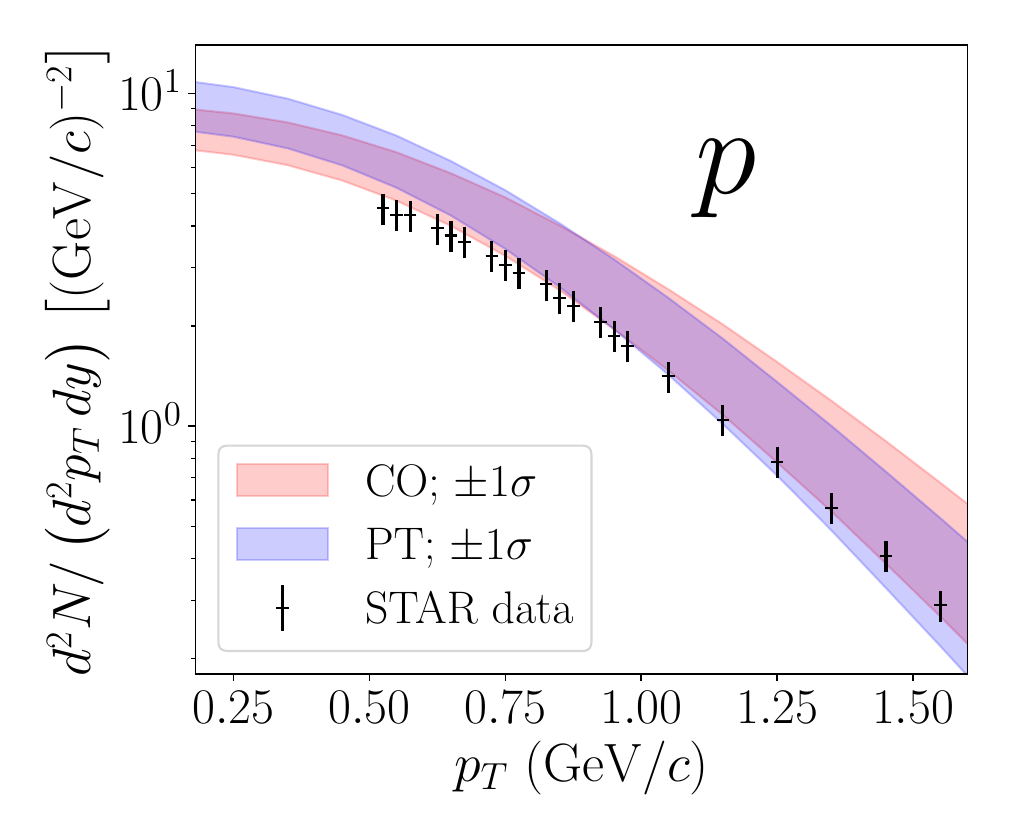}
\includegraphics[width=0.32\textwidth,height=0.22\textheight]{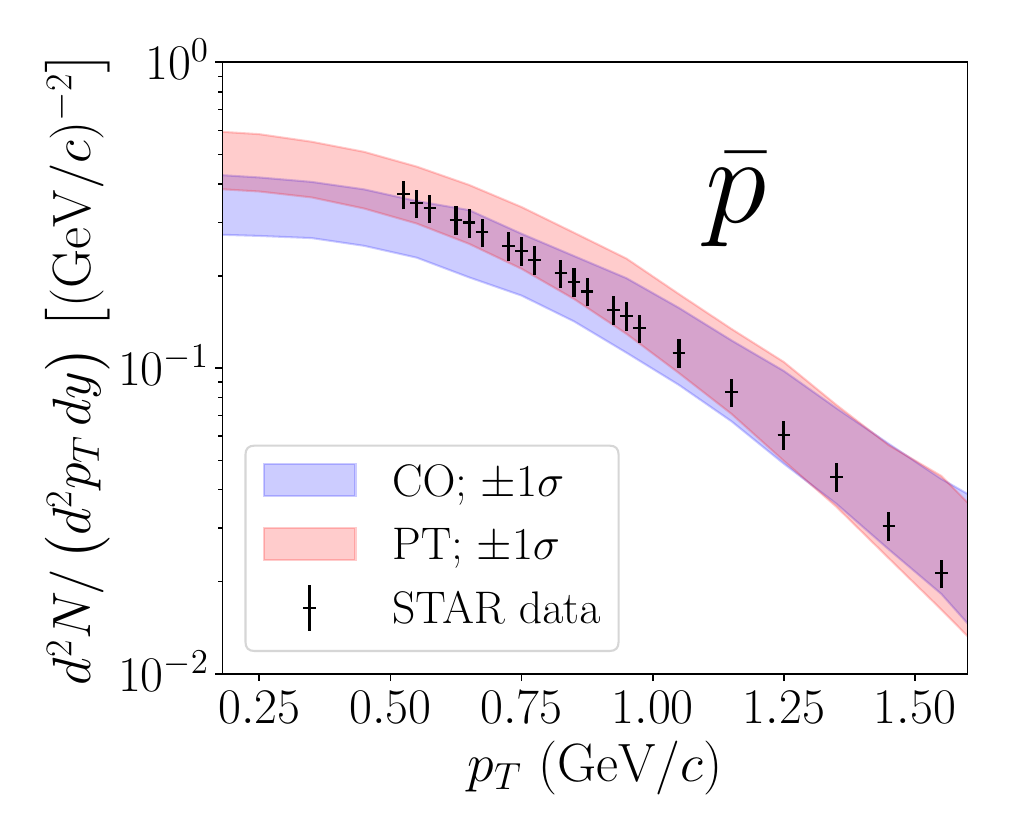}

\caption{\label{fig:avaraged_plots} 
Midrapidity ($|y|<0.1$) transverse momentum spectra 
$d^2N/\left(2\pi p_T dp_T dy\right)$ for $\pi^-$, $p$, and $\bar{p}$ 
(from left to right), averaged over 750 simulation sets with random parameters within the ranges in Table~\ref{tab:latinCube}. Shaded regions show one standard deviation from the mean. STAR Collaboration data, used for comparison, correspond to Au+Au collisions at $\sqrt{s_{NN}}=14.5$~GeV in the 20--30\% centrality class~\cite{STAR:2017sal, STAR:2019vcp}.}

\end{figure*}

For each parameter set, we select 40 UrQMD initial conditions corresponding to the $20$–$30\%$ centrality class. To improve statistical accuracy, 200 afterburner events are simulated on the particlization hypersurface for each initial condition, yielding a total of 8,000 events per parameter set. Transverse momentum spectra for pions, kaons, and (anti)protons are then constructed. Fig.~\ref{fig:avaraged_plots} shows the spectra at midrapidity, averaged over 750 parameter sets, with a one-standard-deviation range for both equations of state. On average, midrapidity particle yields are higher for the softer phase-transition equation of state, but this scenario also produces weaker transverse flow, reflected in the high-$p_T$ spectra. The mean parameter values listed in Table~\ref{tab:latinCube} do not correspond to best-fit values for experimental data; consequently, the data may lie outside the one-standard-deviation range from the mean values, as shown by the blue and red shaded bands in Fig.~\ref{fig:avaraged_plots}. This figure also indicates that antiprotons at low $p_T$ are the most sensitive to changes in the equation of state.

Further, for simplicity, we consider only the following $p_t$ bins: $0.15\pm 0.05$, $0.55\pm 0.05$, and $1.05\pm 0.05$~GeV/$c$.  For quantitative analysis, we use the Pearson correlation coefficient between the spectra values $Y_i$ and the model parameters $X_j$:

\begin{equation}
P(X_j, Y_i) = \frac{\text{Cov}(X_j, Y_i)}{\sqrt{\text{Var}(X_j) \text{Var}(Y_i)}}.
\end{equation}

Here, $\text{Cov}$ is the covariance matrix, defined as
\begin{equation}
\text{Cov}(X_j, Y_i) = \frac{1}{N-1}\sum_{k=1}^{N} \left(X_{jk} - \mu^{x}_{j}\right)\left(Y_{ik} - \mu^{y}_{i}\right),
\end{equation}
where the sum is taken over $N=750$ simulation sets, and $\mu^{x}_{j}$ and $\mu^{y}_{i}$ are the mean values of the $j^{\text{th}}$ parameter and the $i^{\text{th}}$ observable, respectively. Finally, $\text{Var}(X_j)$ and $\text{Var}(Y_i)$ are the diagonal elements of the self-covariance matrices $\text{Cov}(X_j, X_j)$ and $\text{Cov}(Y_i, Y_i)$, respectively.

The sign of the correlation matrix element indicates whether the parameter and observable are correlated ($+$) or anticorrelated\footnote{An increase in the parameter value leads to a decrease in the observable value.} ($-$), while the absolute value reflects the strength of the correlation, with ``0'' representing no correlation and ``$\pm$1'' corresponding to a strong (anti) correlation.

\begin{figure*}
\centering

\includegraphics[width=0.22\textwidth]{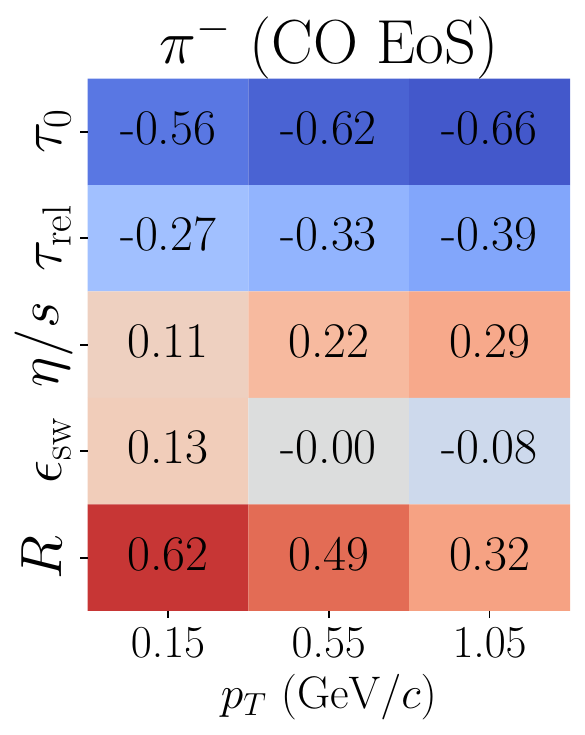}
\includegraphics[width=0.22\textwidth]{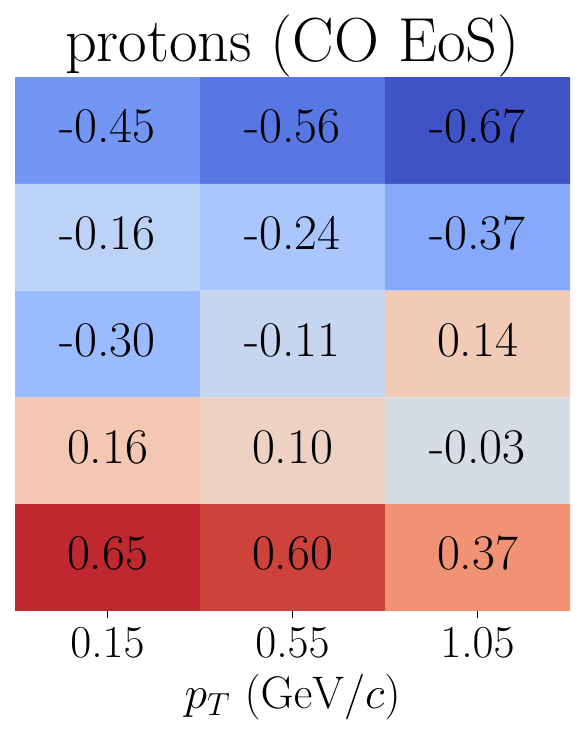}
\includegraphics[width=0.22\textwidth]{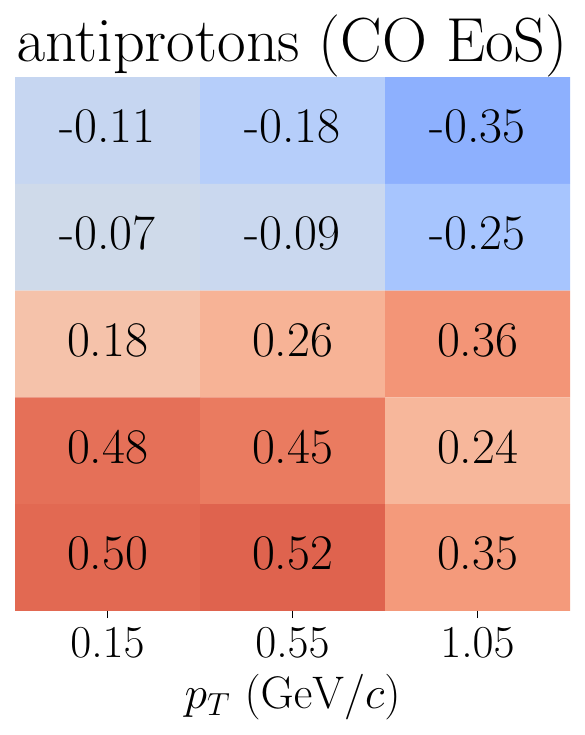}
\includegraphics[width=0.22\textwidth]{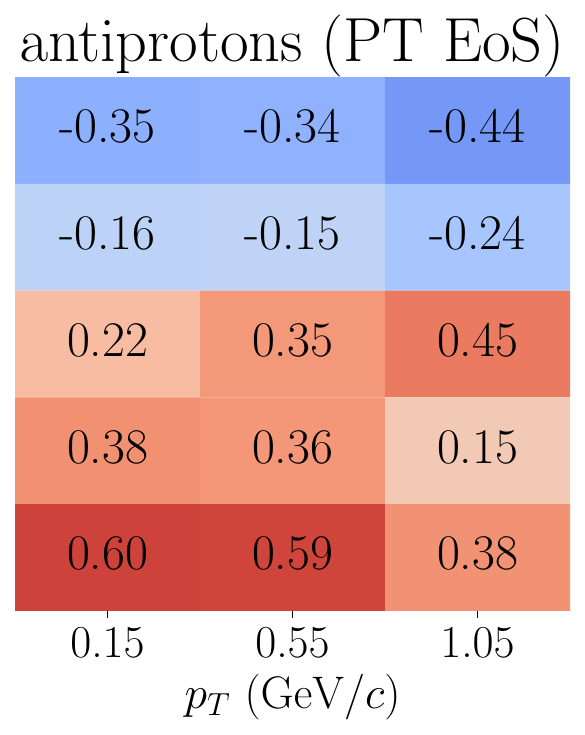}
\includegraphics[width=0.07\textwidth,height=0.2\textheight]{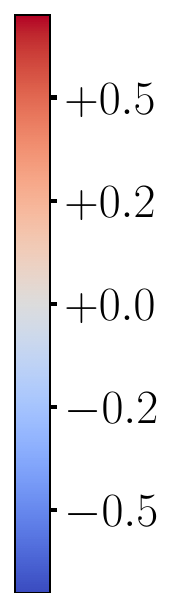}

\caption{\label{fig:correlatins} Correlation coefficients between the five model parameters and the transverse momentum spectra at low, intermediate, and high $p_{\text{T}}$ for $\pi^-$, $p$, and $\bar{p}$ in $20-30\%$ centrality Au+Au collisions at $\sqrt{s_{NN}}=14.5$~GeV. The results for $\pi^{-}$ mesons and protons (shown in the two tables on the left) are very similar for both equations of state; therefore, only the crossover case is shown. For kaons, the correlation matrices are similar to those of pions. For antiprotons (shown in the two plots on the right), we present the results for both scenarios.}
\end{figure*}

We investigate correlation matrices for six particle species: $\pi^{\pm}$, $K^{\pm}$, $p$, and $\bar{p}$, using both equations of state. However, since the results for pions and kaons are very similar, in Figure~\ref{fig:correlatins} we present only the correlations for negatively charged pions, protons, and antiprotons. Additionally, we show correlations only for antiprotons for the soft EoS with a first-order phase transition, as the results for all other cases are qualitatively and quantitatively similar to those of the crossover scenario.

One key conclusion from Fig.~\ref{fig:correlatins} is that the commonly used smoothing procedure for initial conditions, encoded in the parameter $R$, strongly influences hydrodynamic simulations and limits our ability to extract properties of the matter created in heavy-ion collisions. A second observation is that the parameter $\tau_{\rm rel}$ behaves similarly to $\tau_0$, although its impact on the observables is less pronounced. We also find that, unlike other hadron species, antiproton yields are particularly sensitive to the switching energy density $\epsilon_{\text{sw}}$, which marks the transition from hydrodynamics to transport. This sensitivity, also demonstrated in Ref.~\cite{Monnai:2019hkn}, can be attributed to the suppression of (anti)baryon yields in transport models such as UrQMD \cite{Bass:1998ca, Bleicher:1999xi} or SMASH \cite{Petersen:2018jag}, which are employed as afterburners in the final stage of the system's evolution. This issue has been addressed in numerous studies, to which we refer the reader \cite{Garcia-Montero:2021haa, Savchuk:2021aog, Becattini:2012sq}.

When comparing the correlation matrices for antiprotons, we observe that in the case of a softer equation of state (PT EoS), the parameters defining the pre-thermal evolution (i.e., $\tau_0$, $\tau_{\text{rel}}$, $R$) play a more significant role than in the case of the crossover equation (CO EoS). This is reflected in the larger absolute values of the corresponding correlation coefficients. In contrast, the particlization energy density ($\epsilon_{\text{sw}}$), which defines the late-stage evolution, becomes more influential when a stiffer equation of state is employed. Finally, we note that Pearson correlations capture only linear relationships. The results may depend on collision energy, particularly at low energies where the system spends longer in the transition region of the QCD phase diagram and the differences between equations of state are more pronounced.

\subsubsection{Model finetuning}
In this study, we calibrate the model's free parameters using only the particle spectra. As illustrated in Fig.~\ref{fig:correlatins}, the starting time of the thermalization stage, $\tau_0$, is one of the most important parameters. Together with the relaxation time $\tau_{\rm rel}$, they are constrained by the pion multiplicities and the slope of the transverse momentum spectra. In principle, these timescales can depend on the collision centrality. However, our numerical analysis shows that if $\tau_0$ is calibrated separately for different centrality classes (between $0$--$5$\% and $30$--$40$\%), the optimal values differ by no more than $0.1$~fm/$c$ at $\sqrt{s_{_{\rm NN}}}=39$~GeV and by about $0.2$~fm/$c$ at $\sqrt{s_{_{\rm NN}}}=7.7$~GeV for both equations of state. Therefore, to reduce the number of free parameters, we present the results obtained for the $20$--$30$\% centrality class, where the system is sufficiently large yet exhibits transverse asymmetry due to the nonzero impact parameter.

Since low-$p_T$ spectra exhibit only a weak dependence on $\eta/s$ \cite{Song:2007ux}, we fix this parameter at a typical value of $\eta/s = 0.08$ \cite{Romatschke:2007mq}, deferring its detailed tuning to future studies with a more comprehensive analysis. Because antiproton spectra are particularly sensitive to the switching energy density $\epsilon_{\rm sw}$, we use experimental antiproton data to constrain this parameter. Although the optimal values of $\epsilon_{\rm sw}$ differ for the two equations of state, we do not vary them when changing the collision energy $\sqrt{s_{NN}}$. 

The smoothing parameter $R$ appears in various forms across most dynamical models with event-by-event fluctuating initial conditions, yet its value remains poorly constrained. For example, Ref.~\cite{Akamatsu:2018olk} uses $R = 0.5 \ \text{fm}$, Ref.~\cite{Inghirami:2022afu} adopts $1 \ \text{fm}$, and Ref.~\cite{Schafer:2021csj} treats it as a free parameter. In iHKM, smoothing is further influenced by the thermalization stage, and there is no strong evidence to vary $R$ across different collision energies. Therefore, we set $R$ to the smallest reasonable value of $0.5 \ \text{fm}$. Smaller values result in overly spiky initial distributions, which can destabilize the early hydrodynamic evolution.

\begin{figure}
\includegraphics[width=0.5\textwidth]{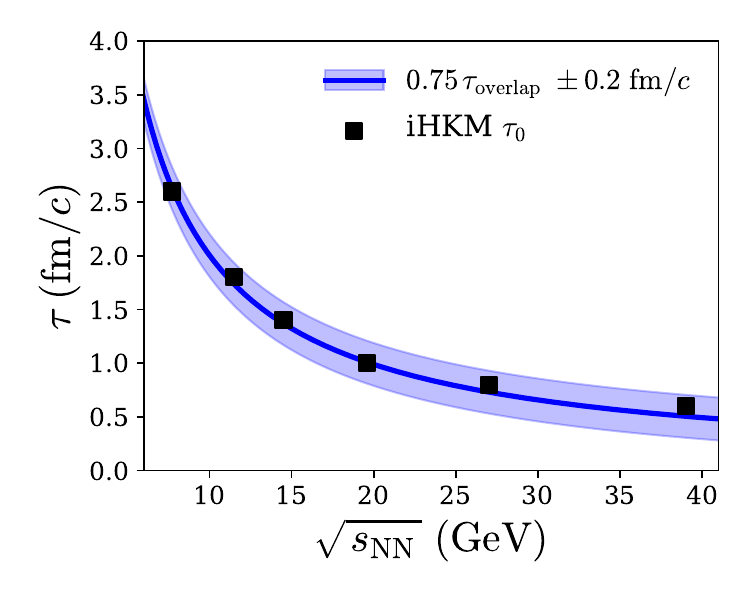}
\caption{\label{fig:tau0} 
Inferred values of the $\tau_0$ parameter in iHKM as a function of collision energy $\sqrt{s_{NN}}$ (black markers). 
The blue line shows the nuclei overlapping time from Eq.~(\ref{eq:overlapTime}), scaled by a factor of $0.75$, and the shaded region represents the estimated uncertainty of $\tau_0$.}
\end{figure}

The final iHKM parameters for different RHIC BES energies are summarized in Table~\ref{tab:params}. Figure~\ref{fig:tau0} shows the dependence of the extracted $\tau_0$ values on the collision energy $\sqrt{s_{NN}}$. We find that $\tau_0$ remains close to the nuclei overlap time, $\tau_{\text{overlap}}$, calculated from Eq.~(\ref{eq:overlapTime}) and scaled by a factor of 0.75. The deviations do not exceed $\pm 0.2$~fm/$c$, which corresponds to the uncertainty in the calibration of $\tau_0$. It should be noted, however, that this overlap time corresponds to zero impact parameter, while our analysis indicates nearly the same $\tau_0$ values even for $30$--$40$\% non-central collisions, where the nuclei overlap for a shorter duration. This behavior can be attributed to the assumption of uniform thermalization across the system as a function of proper time in our model. In a more realistic scenario, the thermalization rate could depend on the local energy density; however, such an approach would significantly complicate the hydrodynamic equations. Furthermore, the available experimental data for particle spectra are restricted to a narrow midrapidity window, $\left|y\right| < 0.1$, which further limits the precision of the parameter extraction.

Another outcome of the model calibration is an estimate of the duration of the thermalization stage. As shown in Table~\ref{tab:params}, the characteristic scale of
$\tau_{\text{rel}} = \tau_{\text{th}} - \tau_0$
is of the order of 1~fm/$c$, and no correlation with the collision energy is observed. Likewise, the end of the hydrodynamic stage, characterized by the maximum proper time on the hypersurface of constant energy density $\epsilon_{\text{sw}}$ (hereafter denoted as $\tau_{\text{sw}}^{\max}$ for convenience), exhibits only a weak dependence on the collision energy. For example, in the most central collisions, when decreasing the collision energy from 39~GeV to 7.7~GeV, $\tau_{\text{sw}}^{\max}$ increases only from 9.2~fm/$c$ to 9.8~fm/$c$. For noncentral collisions, this trend is slightly more pronounced. In the 20–30\% centrality class, the corresponding times are 8.8~fm/$c$ and 11.5~fm/$c$. Notably, in all these cases, the hydrodynamic stage ends considerably later than the thermalization time $\tau_{\text{th}}$. This indicates that, at the RHIC BES energies considered in this paper, most of the system undergoes thermalization and subsequently evolves through a relatively long hydrodynamic stage. Consequently, the system is largely thermalized, except for the peripheral regions of the fireball with low initial energy density.

Extrapolating these trends to lower collision energies, a qualitative change in the system’s evolution can be anticipated. If $\tau_0$ continues to scale with the nuclear overlap time $\tau_{\text{overlap}}$, which increases rapidly in the few-GeV regime, while the end of the hydrodynamic stage $\tau_{\text{sw}}^{\max}$ remains only weakly dependent on energy, full hydrodynamization is not achieved for most of the system. Consequently, particle spectra are expected to contain a substantial non-thermal component. Quantitative estimates indicate that at energies around 4.5~GeV (accessible at STAR FXT~\cite{STAR:2020dav} and FAIR CBM~\cite{Agarwal:2022ydl}), where $\tau_{\text{overlap}} \approx 7.5$~fm/$c$, this non-thermal contribution (see Eq.~\ref{eq:dist_function}) is already significant. At lower energies, near 3~GeV (studied also by HADES~\cite{HADES:2020wpc}), it is expected to dominate. These low-energy regimes will be the focus of our future studies.

\begin{table}[]
\caption{\label{tab:params}
Optimized iHKM parameters that provide the best description of transverse momentum spectra in Au+Au collisions at $\sqrt{s_{NN}} =7.7, 11.5, 14.5, 19.6, 27.0$, and $39.0$~GeV. The remaining parameters are fixed at $R = 0.5$~fm and $\eta/s = 0.08$.}

\begin{ruledtabular}
\begin{tabular}{@{}ccccc@{}}
$\sqrt{s_{NN}} $  (GeV)& EoS & $\tau_{0}$ (fm/$c$) & $\tau_{\text{th}}$ (fm/$c$) & $\epsilon_{\text{sw}}$ (GeV/fm$^3$) \\ \hline
7.7  & PT & 2.7 & 3.3 & 0.35 \\
7.7  & CO & 2.5 & 3.3 & 0.50 \\ 
11.5 & PT & 2.0 & 2.6 & 0.35 \\
11.5 & CO & 1.8 & 2.6 & 0.50 \\ 
14.5 & PT & 1.4 & 2.0 & 0.35 \\
14.5 & CO & 1.3 & 2.3 & 0.50 \\ 
19.6 & PT & 1.0 & 1.6 & 0.35 \\ 
19.6 & CO & 1.0 & 1.6 & 0.50 \\
27.0 & PT & 0.8 & 1.4 & 0.35 \\ 
27.0 & CO & 0.8 & 1.4 & 0.50 \\ 
39.0 & PT & 0.6 & 1.2 & 0.35 \\ 
39.0 & CO & 0.6 & 1.6 & 0.50 \\ 
\end{tabular}
\end{ruledtabular}
\end{table}

\subsubsection{Momentum spectra}
\begin{figure*}
    \centering
    \begin{subfigure}{0.5\textwidth}
        \centering
        \includegraphics[width=\textwidth]{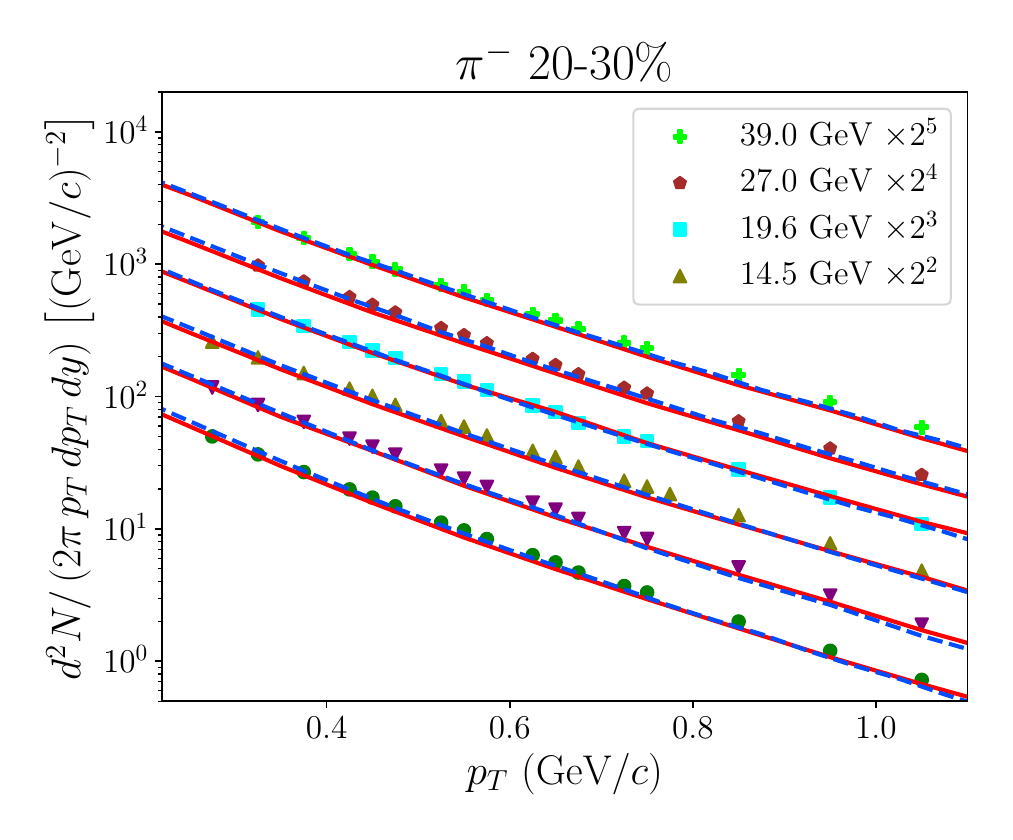}
    \end{subfigure}%
    \begin{subfigure}{0.5\textwidth}
        \centering
        \includegraphics[width=\textwidth]{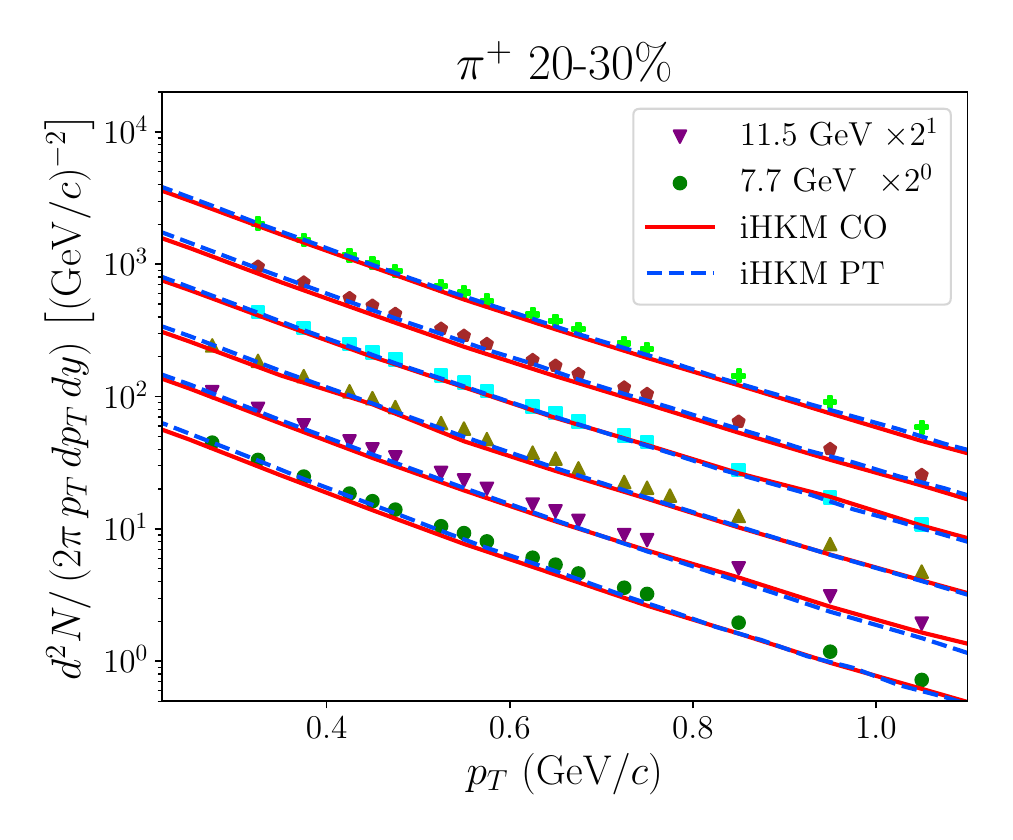}
    \end{subfigure}%
    \hfill
    
    \begin{subfigure}{0.5\textwidth}
        \centering
        \includegraphics[width=\textwidth]{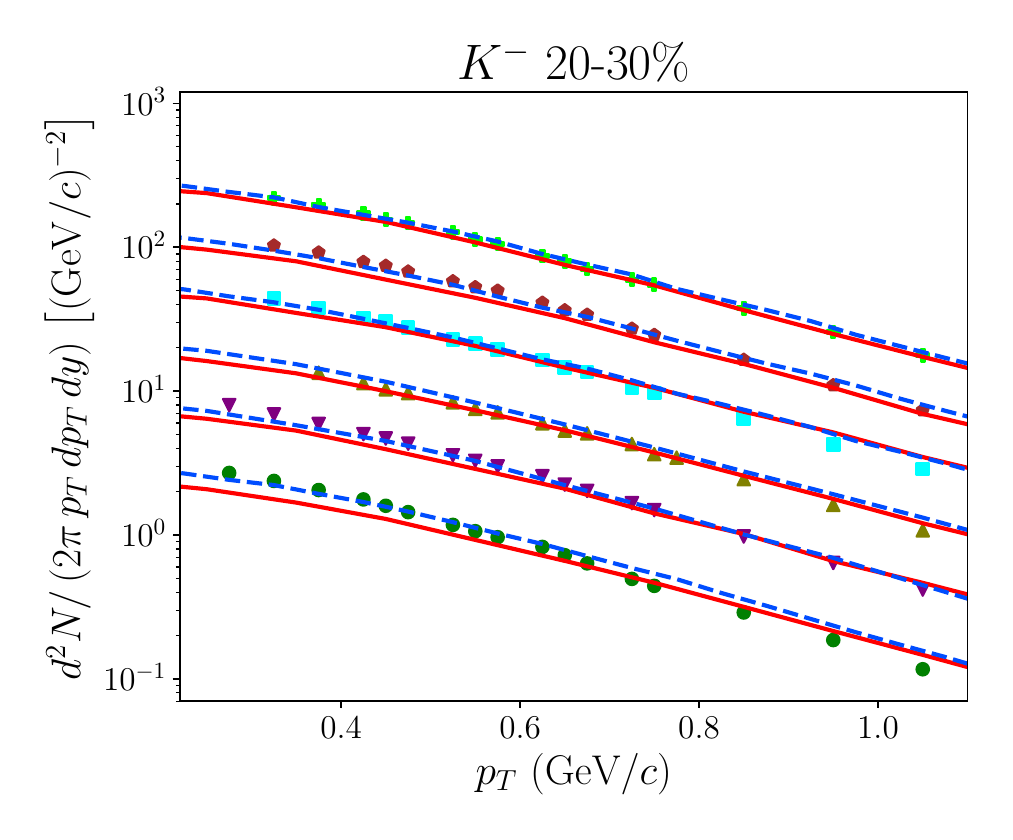}
    \end{subfigure}%
    \begin{subfigure}{0.5\textwidth}
        \centering
        \includegraphics[width=\textwidth]{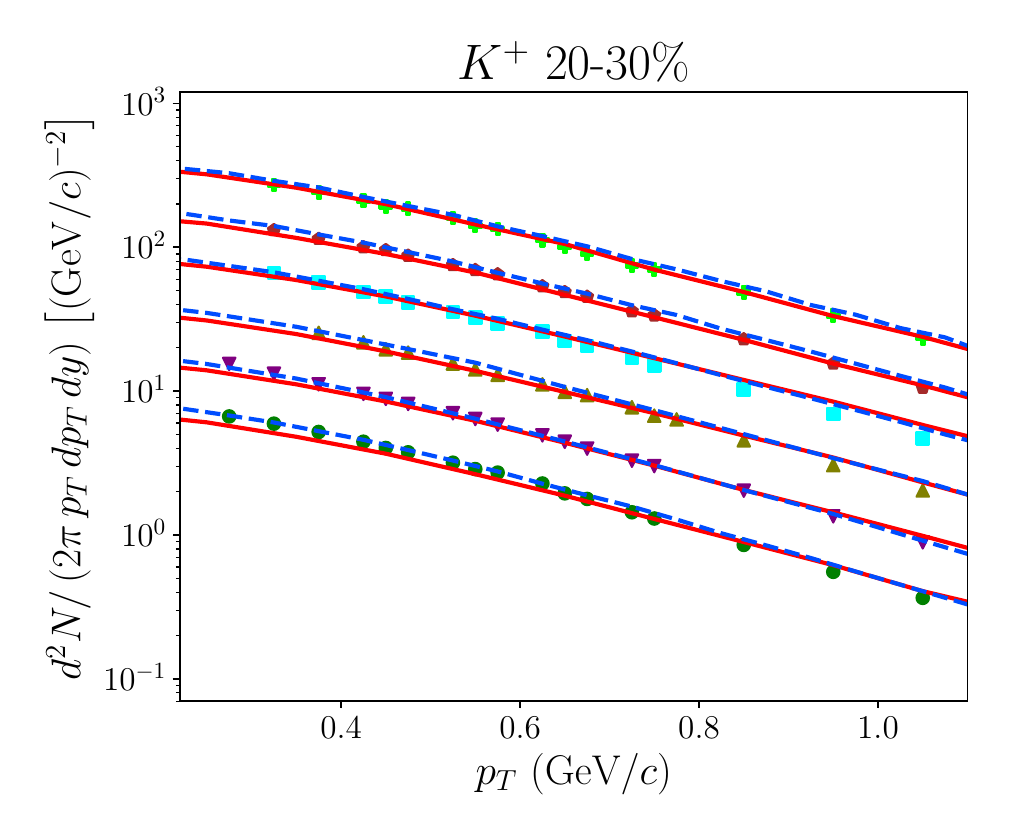}
    \end{subfigure}%
    \hfill
    
    \begin{subfigure}{0.5\textwidth}
        \centering
        \includegraphics[width=\textwidth]{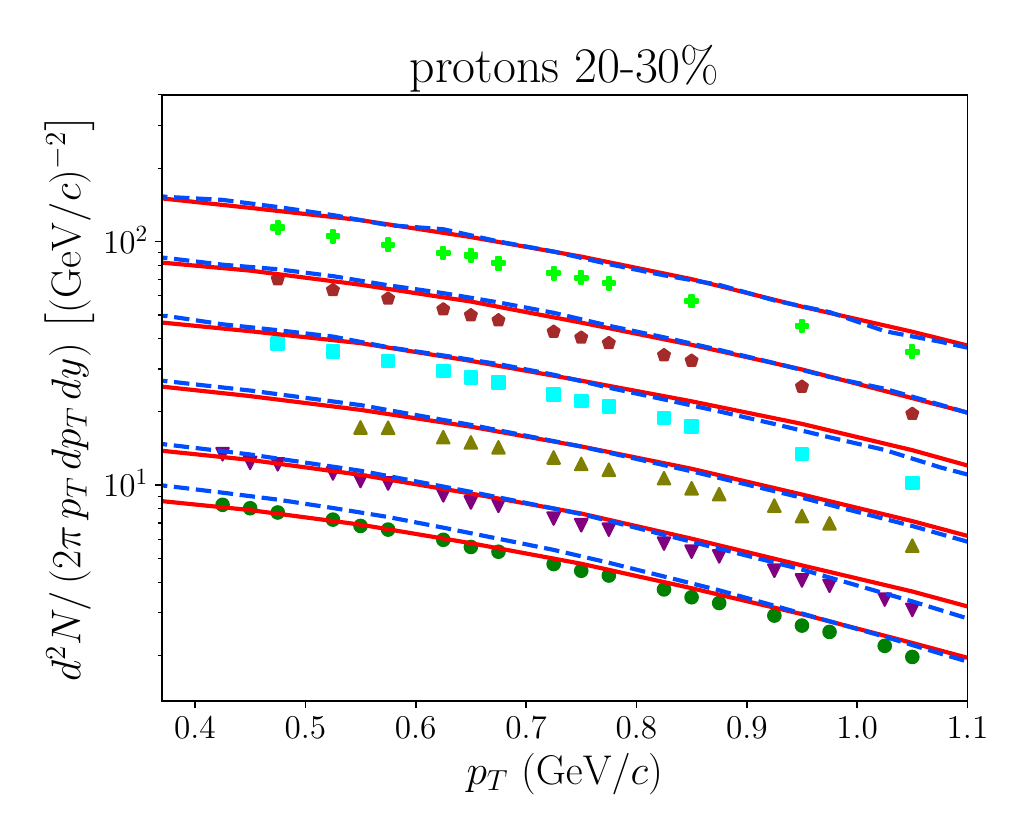}
    \end{subfigure}%
    \hfill
    \begin{subfigure}{0.5\textwidth}
        \centering
        \includegraphics[width=\textwidth]{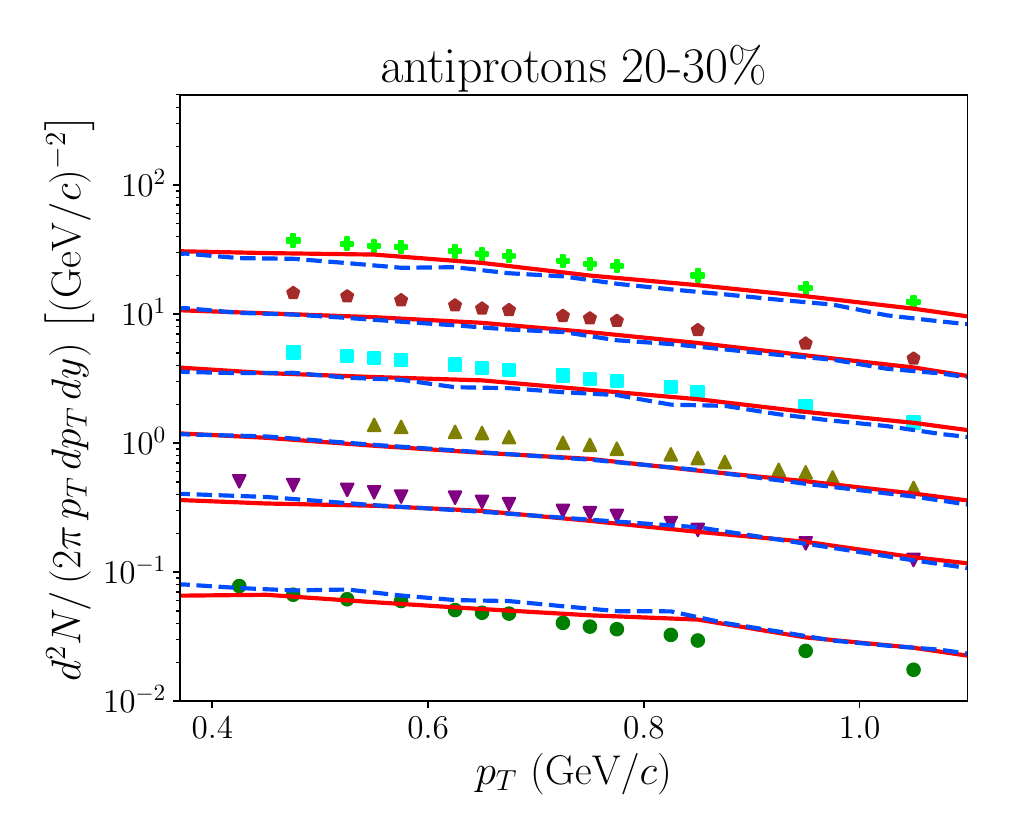}
    \end{subfigure}%
    
    \caption{\label{fig:2030} Comparison of iHKM results with STAR data~\cite{STAR:2017sal, STAR:2019vcp} for $\pi^{\pm}$, $K^{\pm}$, $p$, and $\bar{p}$ transverse momentum spectra in the 20--30\% centrality class at midrapidity ($|y| < 0.1$). The pion spectra were used to calibrate the initial time parameters $\tau_0$ and $\tau_{\text{th}}$.}
    
\end{figure*}

\begin{figure*}
    \centering
    \begin{subfigure}{0.5\textwidth}
        \centering
        \includegraphics[width=\textwidth]{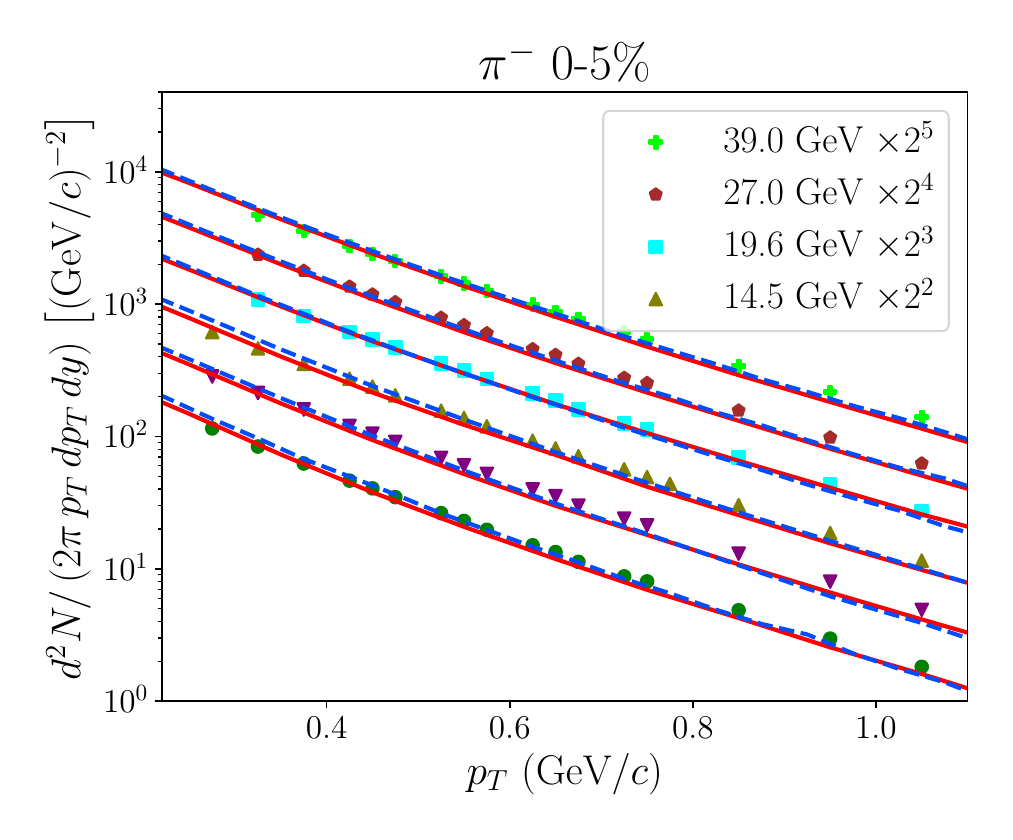}
    \end{subfigure}%
    \begin{subfigure}{0.5\textwidth}
        \centering
        \includegraphics[width=\textwidth]{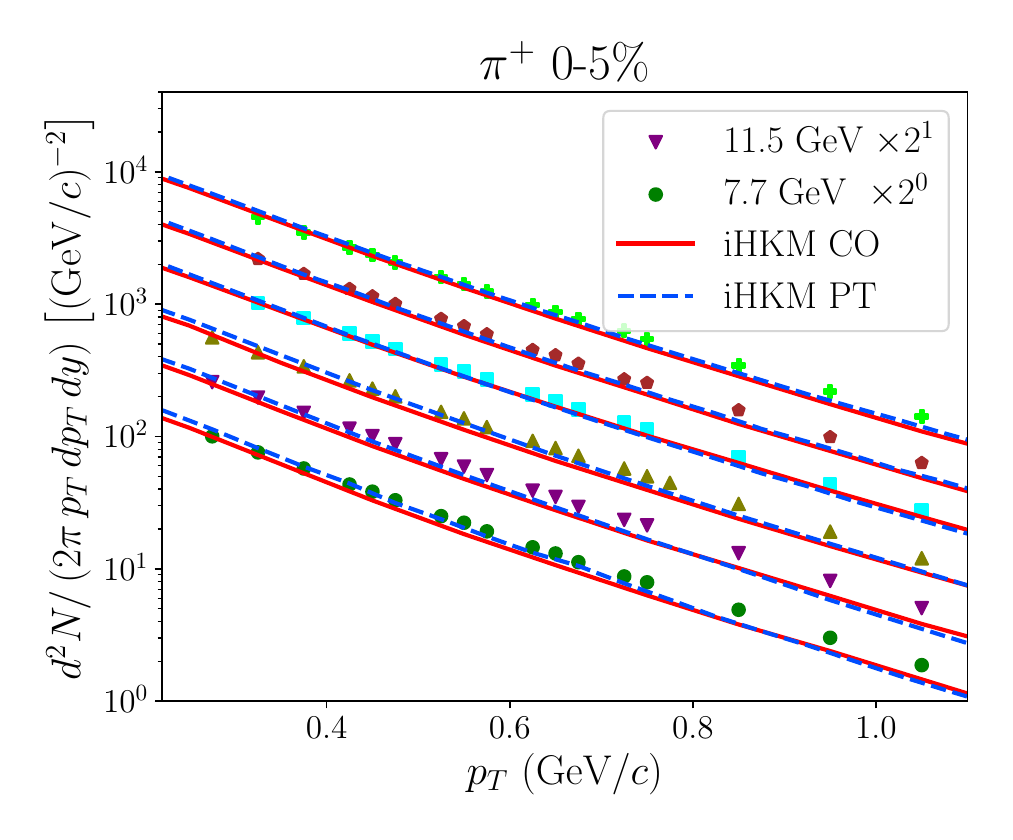}
    \end{subfigure}%

    \begin{subfigure}{0.5\textwidth}
        \centering
        \includegraphics[width=\textwidth]{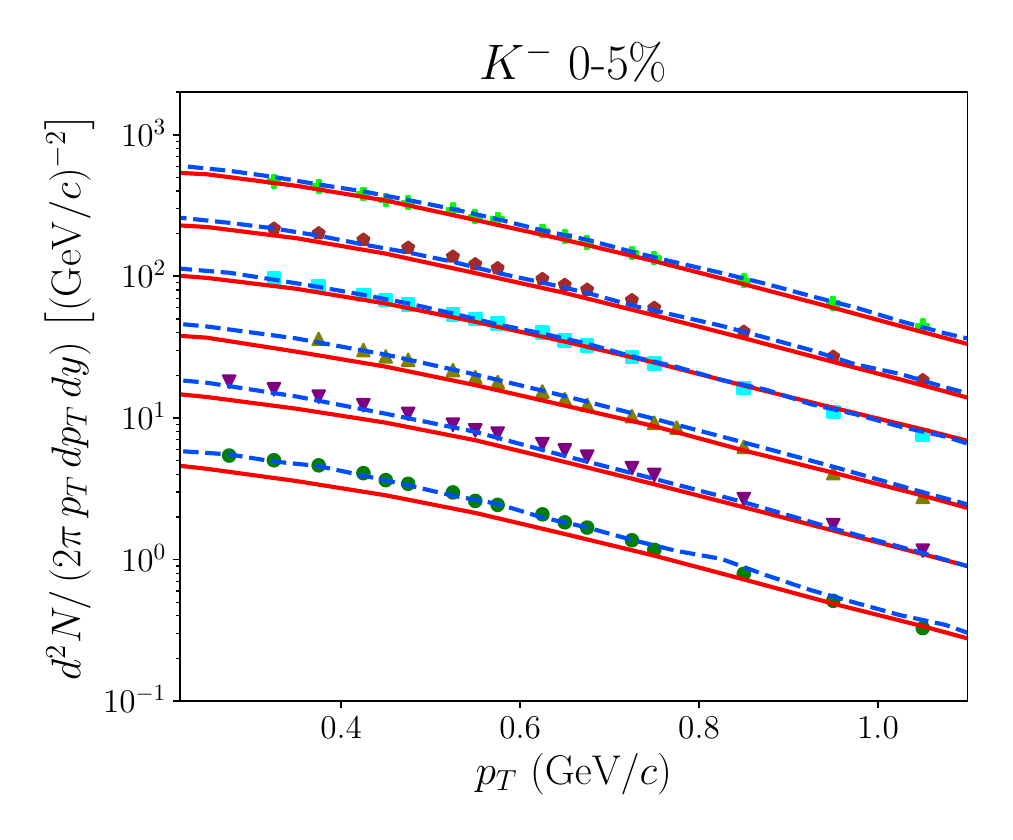}
    \end{subfigure}%
    \begin{subfigure}{0.5\textwidth}
        \centering
        \includegraphics[width=\textwidth]{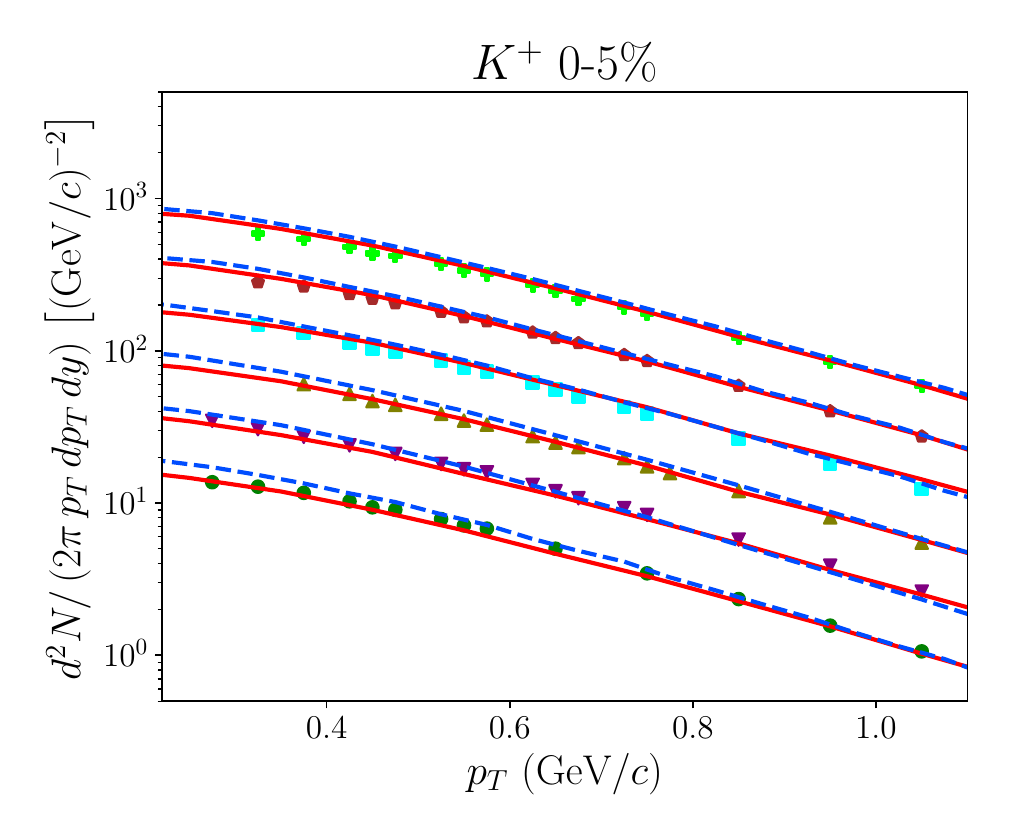}
    \end{subfigure}%

    \hfill
    
    \begin{subfigure}{0.5\textwidth}
        \centering
        \includegraphics[width=\textwidth]{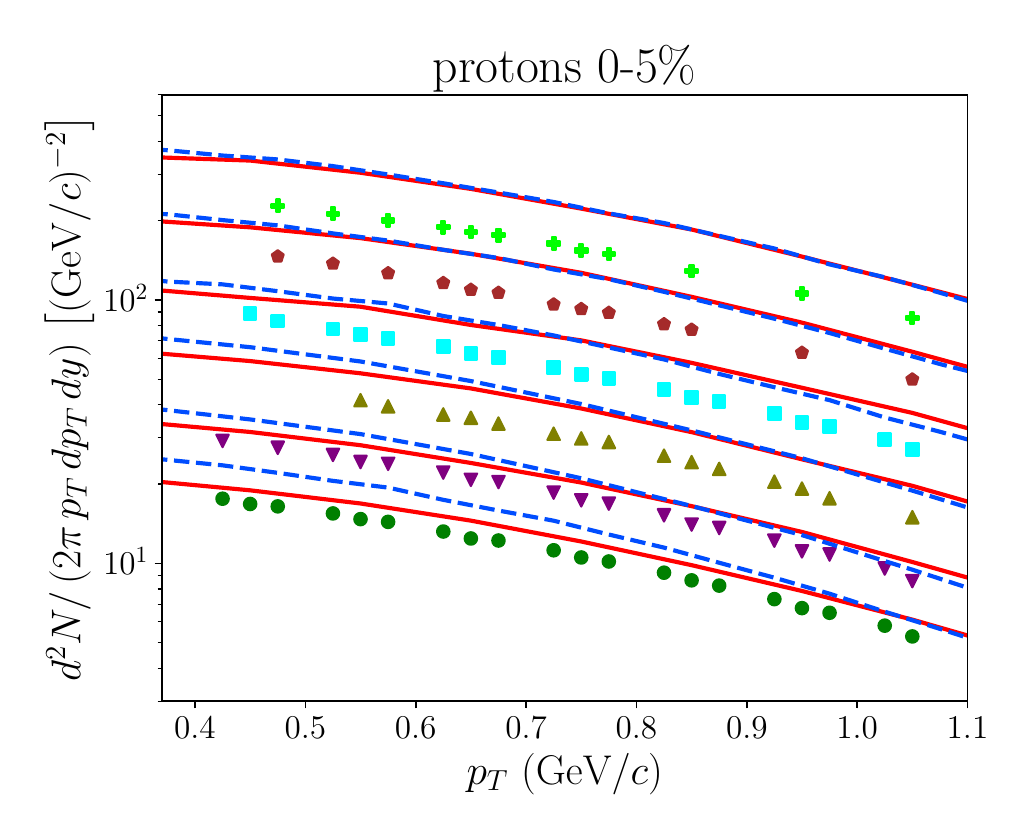}
    \end{subfigure}%
    \begin{subfigure}{0.5\textwidth}
        \centering
        \includegraphics[width=\textwidth]{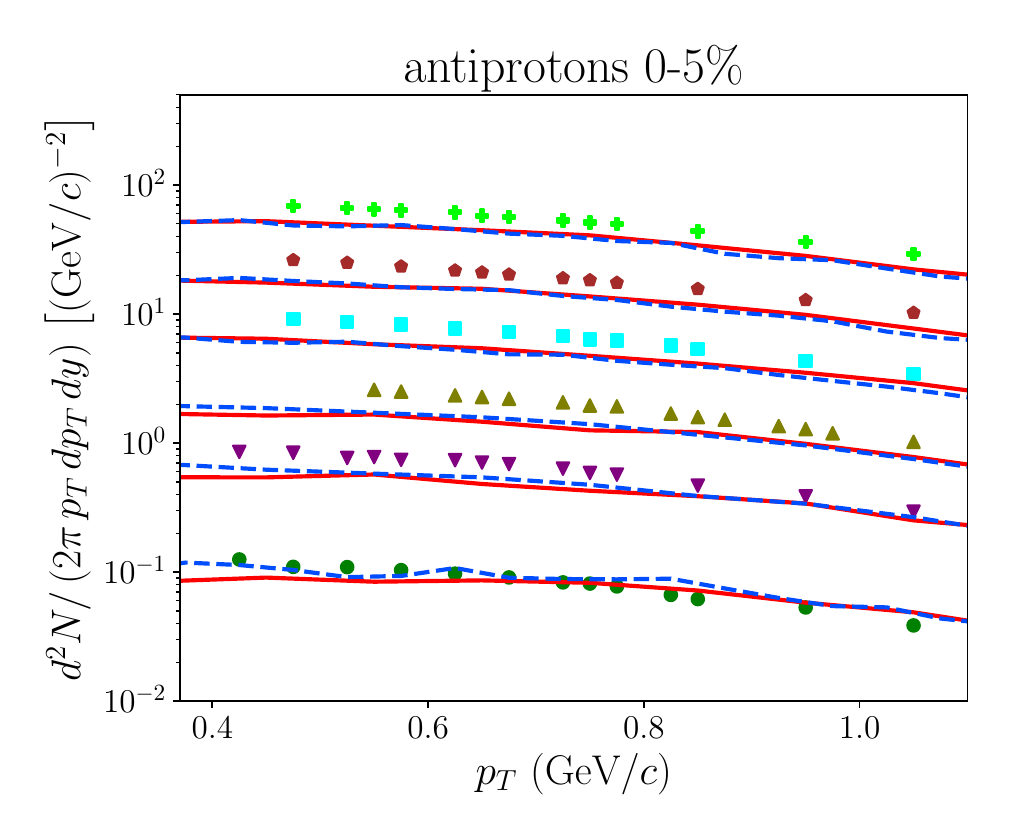}
    \end{subfigure}%
    
    \caption{\label{fig:0005} Comparison of iHKM results with STAR data~\cite{STAR:2017sal, STAR:2019vcp} for $\pi^{\pm}$, $K^{\pm}$, $p$, and $\bar{p}$ transverse momentum spectra in the 0--5\% centrality class at midrapidity ($|y| < 0.1$). Model parameters are the same as for the 20--30\% centrality class.}
\end{figure*}

We now present the iHKM results for transverse momentum spectra in Au+Au collisions at energies of 7.7, 11.5, 14.5, 19.6, 27, and 39~GeV, for two centrality classes: 20–30\% and 0–5\%, as shown in Figs.~\ref{fig:2030} and \ref{fig:0005}. Analysis of these spectra reveals several key trends. First, the differences between the results obtained with the two equations of state are relatively small and become negligible at higher collision energies. This behavior is also reflected in the model parameters listed in Table~\ref{tab:params}. The similarity arises from the nearly identical description of the QGP phase in both equations of state, with the main differences occurring near the transition to the hadron resonance gas. At higher collision energies, the rapid expansion of the system makes this transition occur too quickly for the differences to significantly affect the spectra, whereas at lower energies the differences are more pronounced.

Second, at midrapidity, the model consistently overestimates proton yields and underestimates antiproton yields across a wide range of transverse momenta. In extreme cases, the deviation between the model and experimental data can reach up to 40\% at small transverse momenta $p_T$. Similar discrepancies in proton spectra have been reported in several other models~\cite{Schafer:2021csj, Gopal:2024qia, Cimerman:2023hjw, Stefaniak:2022pxc}, including those employing cascade-based initial conditions, such as SMASH~\cite{SMASH:2016zqf} and UrQMD~\cite{Bleicher:1999xi}. 

These discrepancies may stem from an imbalance between baryon creation and annihilation processes during the hadronic afterburner stage, or from an inaccurate baryon density distribution on the particlization hypersurface, leading to the over- or underproduction of baryons. In turn, the baryon densities on the particlization hypersurface are determined by the initial conditions (especially the baryon current) and the system's subsequent evolution during the pre-equilibrium and hydrodynamic stages. However, the influence of these factors is beyond the scope of the present study. 

It is also worth noting that in Ref.~\cite{Jahan:2024wpj}, the proton spectra show significantly better agreement with the experimental data at $\sqrt{s_{NN}} = 19.6$~GeV. This improvement, however, may result from the use of more flexible initial conditions, which introduce additional free parameters.

\section{Summary}

In this paper, we employ a recently developed, extended version of the iHKM model to describe the transverse momentum spectra of light hadrons in the Beam Energy Scan program at the Relativistic Heavy Ion Collider. Our primary focus is to explore the relationship between the model's free parameters and experimental observables. In particular, we investigate how variations in the thermalization time scales and the switching conditions between the hydrodynamic and hadronic stages influence particle spectra and transverse momentum distributions. Simulations are performed using the two distinct equations of state: one featuring a crossover transition in baryon-rich matter between quark-gluon plasma and hadron gas, and the other incorporating a first-order phase transition.

Our analysis reveals that the duration of the thermalization stage is approximately 1~fm/$c$ across all RHIC BES energies in the range $\sqrt{s_{NN}} = 7.7$--$39$~GeV. However, thermalization begins later at lower collision energies. Specifically, our numerical results indicate that thermalization commences slightly before the two colliding nuclei fully overlap, assuming they propagate with their initial rapidities; this corresponds to $\tau_0 \approx 0.75\, \tau_{\text{overlap}}$ for both considered equations of state. 

We also find that both equations of state provide comparably accurate descriptions of the transverse momentum spectra, once the relaxation timescale is appropriately adjusted. At higher collision energies within the BES range, differences between the two scenarios in the spectra become negligible. A detailed analysis of other bulk observables is therefore required.

The most significant differences between the two equations of state appear at $\sqrt{s_{NN}} = 7.7$~GeV, particularly in the proton and kaon yields. These discrepancies likely arise from the distinct trajectories the systems follow on the $T$--$\mu_B$ phase diagram during their evolution. The interplay between the phase transition line and the choice of switching energy density, $\epsilon_{\text{sw}}$, for the hadronic afterburner determines the system’s configuration at chemical freeze-out. A more comprehensive study of additional bulk observables, specifically elliptic flow and femtoscopy analysis, within the iHKM framework is planned for future work.

\begin{acknowledgments}
H.Z.'s work was supported by the Grant of National Science Centre, Poland, No: 2021/41/B/ST2/02409 and 2020/38/E/ST2/00019. The research of Yu. S. was partially funded by IDUB-POB projects granted by WUT (Excellence Initiative: Research University (ID-UB)). The research of
M.A. was funded by the National Academy of Sciences of Ukraine through the grant for young scientist research groups. M.A. also gratefully acknowledges support from the Simons
Foundation (Grant No. SFI-PD-Ukraine-00014578).
\end{acknowledgments}










\bibliographystyle{apsrev4-2} 
\bibliography{apssamp}

\end{document}